\documentclass[12pt]{article}
\begin{document}
\newcommand{\bq}{\begin{equation}}
\newcommand{\eq}{\end{equation}}
\newcommand{\bqa}{\begin{eqnarray}}
\newcommand{\eqa}{\end{eqnarray}}
\newcommand{\nl}{\nonumber \\}
\newcommand{\suml}{\sum\limits}
\newcommand{\f}{\varphi}
\newcommand{\eqn}[1]{Eq.(\ref{#1})}
\newcommand{\bpsi}{\bar{\psi}}
\newcommand{\bx}{\bar{x}}
\newcommand{\bchi}{\bar{\chi}}
\newcommand{\hphi}{\hat{\phi}}

\begin{center}
{\bf {\Large
Counting tree diagrams: asymptotic results for QCD-like theories}\\
\vspace*{\baselineskip}
P. Draggiotis\footnote{{\tt petros@sci.kun.nl}}
 and R. Kleiss\footnote{{\tt kleiss@sci.kun.nl}}\\
University of Nijmegen, Nijmegen, the Netherlands}\\
\vspace*{3\baselineskip}
Abstract
\end{center}
We discuss the enumeration of Feynman diagrams at tree order for processes
with external lines of different types. We show how
this can be done by iterating algebraic Schwinger-Dyson equations.
Asymptotic estimates for
very many external lines are derived. Applications include
QED, QCD and scalar QED, and the asymptotic 
estimates are numerically confronted with the exact results.

\newpage
\section{Introduction}
With the growing complexity of scattering 
amplitudes that are becoming amenable to
calculation, especially with the availability of recursive algorithms
like \cite{BerGie,CarMor}, the question of the number of contributing
Feynman graphs becomes of interest in its own right. As will be seen, already
for tree diagrams their enumeration is a nontrivial question.
In \cite{ArgKlePap} the number of graphs for theories with a single,
self-interacting scalar field was studied, and a method derived to
estimate the asymptotic number of tree graphs for the $1\to n$ amplitude
for large $n$ was described, improving on earlier estimates
\cite{Kui,Vol}.
The precise number of graphs, but not its asymptotic form,
for the case of QCD was discussed also in \cite{Manetal}, albeit
in a manner that does not lead to straightforward numerical results.
It is the aim of the present paper to improve on this situation.
In section 2, we briefly review the single-field case. In section 3, we
discuss the case of `QED', that is, processes involving fermions as well
as bosons with QED-like interactions between them, and also the case where
the bosonic field has a three-boson interaction: for these cases,
almost-exact asymptotic results are easily found. The more complicated
cases of QCD (with additional four-point boson vertices), scalar QED and
`all-out' theories with arbitrary bosonic self-interactions, are discussed
in section 4.

\section{Single-field theories}
The enumeration of tree diagrams in a given theory is simpler than
that of general diagrams with loops, in the sense that for tree
diagrams symmetry factors do not occur: indeed, at present the counting
of higher-order diagrams appears almost hopeless. Since diagrams can be
counted by simply replacing all propagators and vertices by unity, the counting
problem becomes equivalent to solving the Schwinger-Dyson (SD) equation for
a simple zero-dimensional Euclidean theory. To set the stage, let us consider 
such a theory, with a single self-interacting scalar field $\f$ and a
Euclidean action
\bq
S(\f) = {1\over2}\f^2 - V(\phi)\;\;\;,\;\;\;
V(\f) = {1\over3!}\f^3 + {1\over4!}\f^4 +\cdots
\eq
where $V(\phi)$ collects all higher-point interactions according to whether
they are present in the theory or not.
Note that the minus sign and the factorials in front
of the interaction terms ensure that all vertices
are precisely unity. Let us denote by
$a(n)$ the number of tree diagrams entering in the $1\to n$ amplitude,
and by $\phi(z)$ its generating function:
\bq
\phi(z) \equiv \suml_{n\ge1}{z^n\over n!}a(n)\;\;.
\eq
This generating function, then, obeys the algebraic Schwinger-Dyson equation
\bq
\phi(z) = z + V'(\phi(z))\;\;,
\label{basic}
\eq
where double-counting is avoided by the factorials in front of the
terms in the interaction potential: for instance, in pure $\f^3$ theory
the Schwinger-Dyson equation, translated back in terms of $a(n)$, reads
\bq
a(n) = \delta_{n,1} + {1\over2}\suml_{n_{1,2}\ge1}{n!\over n_1!n_2!}a(n_1)a(n_2)
\delta_{n,n_1+n_2}\;\;.
\eq
Obtaining the exact number of graphs for given $n$ is simply a matter of
algebraically iterating \eqn{basic} up to the appropriate order
in powers of $z$, which is a trivial task for any halfway
decent computer algebra system. For the asymptotic result, however, 
we have to employ more. The asymptotic form of $a(n)$ is of course
given by the singularity structure of $\phi(z)$ as a function of $z$.
Now, $\phi(z)$ cannot have poles for finite $z$ if $V(\f)$ is
a finite polynomial, since the SD equation cannot then be satisfied.
The singularities must therefore be branch points. Let us write
$z$ as a function of $\phi$:
\bq
z = F(\phi) \equiv \phi - V'(\phi)\;\;.
\eq
We now look for that value of $\phi$ (and $z$) for which the definition of
$\phi(z)$ becomes ambiguous, {\it i.e.\/} where a branch cut starts.
Such points $\phi_0$ will be situated where $d\phi/dz$ duverges, or
\bq
{dF(\phi)\over d\phi}=0\;\;\;\mbox{at $\phi=\phi_0$}\;\;.
\eq
To each of these roots corresponds a value $z_0=F(\phi_0)$, and that $z_0$
which lies closest to the origin determines the leading asymptotic behaviour.
We can then make an expansion around the appropriate value of $\phi_0$:
\bq
z = F(\phi_0) + {1\over2}F''(\phi_0)(\phi-\phi_0)^2 + \cdots\;\;,
\eq
and we can read off the approximate form of $\phi(z)$ in the neighbourhood
of the singular point:
\bq
\phi(z) \sim \phi_0 - \left(1-{z\over F(\phi_0)}\right)^{1/2}
\sqrt{{-2F(\phi_0)\over F''(\phi_0)}}.
\eq
The expansion of the square-root form,
\bq
\sqrt{1-x} = 1 - \suml_{n\ge1}{(2n-2)!x^n\over n!(n-1)!2^{2n-1}}
\sim \suml_n {x^n\over n^{3/2}\sqrt{4\pi}}\;\;,
\eq
where we indicate its asymptotic form, then tells us that
the asymptotic form of $a(n)$ is given by
\bq
a(n) \sim {n!\over n^{3/2}}C^nD\;\;\;,\;\;\;
C = {1\over F(\phi_0)}\;\;\;,\;\;\;
D=\sqrt{{-F(\phi_0)\over2\pi F''(\phi_0)}}\;\;.
\eq
Two points are in order here. If it should happen that $F''$ vanishes
together with $F'$, we might have to look for a cube-root branch point
rather than a square-root one: we have never yet encountered this case.
Secondly, it is possible, as for instance in pure $\f^4$ theory, that
there are several $z_0$ values equally far from zero. In that case we have
to add the asymptotic contributions from every such point, and this is the
mechanism by which in pure $\f^4$ theory even values of $n$ become
forbidden.

Some results are collected in the the following table,
for theories in which all interactions from $\f^3$ up to and including
$\f^M$ are present.
\begin{center}
\begin{tabular}{|c|c|c|} 
\hline\hline $\vphantom{{A\over A}}$
$M$ &  $C$  &  $D$  \\ \hline
 3   & 2.00000    &  .282095  \\
 4   & 2.50804    &  .191409  \\
 5   & 2.57845    &  .178231  \\
 6   & 2.58755    &  .175794   \\
 7   & 2.58859    &  .175393  \\
 8   & 2.58868    &  .175336  \\
 9   & 2.58869          &  .175331  \\
$\infty$   & 2.58869    &  .175329  \\
\hline\hline
\end{tabular}
\end{center} 
The last theory, with potential $V(\f)=\exp(\f)-\f-1$,
is interesting in that it  establishes an upper bound on the
number of tree graphs in any single-field theory. Equivalently,
it gives the number of graphs arising from an effective action
after tadpole renormalization.

\section{QED-like theories}
We now turn to the more complicated case where fermionic fields are also
present: we then have fermions, antifermions, and bosons. The simplest
case is that of QED with a single fermion type. The action is
then given by
\bq
S(\f,\chi,\bchi) = {1\over2}\f^2 + \bchi\chi - \bchi\chi\f\;\;,
\eq
where $\chi$ denotes the fermion and $\bchi$ the antifermion field.
We now have three amplitudes, depending on the incoming line, and we have
the following generating functions:
\bqa
\phi(z,x,\bx) &=& \suml_{n_{0,1,2}\ge0}
{z^{n_0}x^{n_1}\bx^{n_2}\over n_0!n_1!n_2!}
a(\f\to n_0\f,n_1\chi,n_2\bchi)\;\;,\nl
\psi(z,x,\bx) &=& \suml_{n_{0,1,2}\ge0}
{z^{n_0}x^{n_1}\bx^{n_2}\over n_0!n_1!n_2!}
a(\chi\to n_0\f,n_1\chi,n_2\bchi)\;\;,\nl
\bpsi(z,x,\bx) &=& \suml_{n_{0,1,2}\ge0}
{z^{n_0}x^{n_1}\bx^{n_2}\over n_0!n_1!n_2!}
a(\bchi\to n_0\f,n_1\chi,n_2\bchi)\;\;,
\eqa
and coupled SD equations:
\bq
\phi = z + \psi\bpsi\;\;\;,\;\;\;
\psi = x + \phi\psi\;\;\;,\;\;\;
\bpsi = \bx + \phi\bpsi\;\;.
\eq
These can be readily expressed in $\phi$ alone:
\bq
\psi = {x\over1-\phi}\;\;\;,\;\;\;\bpsi = {\bx\over1-\phi}\;\;,
\eq
and
\bq
\phi = z + {\xi\over(1-\phi)^2}\;\;\;,\;\;\;\xi=x\bx\;\;.
\label{QEDeqn}
\eq
The combination $\xi$ implies, of course, conservation of fermion number,
and we see that it suffices to determine $\phi$ as a function of $z$
and $\xi$, except for processes without any external bosons. Again,
combined expansion in powers of $\xi$ and $z$ is trivial by iterating
\eqn{QEDeqn}. For the asymptotic behaviour we now have to study
a two-variable problem. We do this by momentarily keeping $z$ fixed, and
considering the singularity of $\phi$ in terms of $\xi$:
\bq
\xi = F_z(\phi) = (1-\phi)^2(\phi-z)\;\;\;,\;\;\;
F'_z(\phi) = (1-\phi)(1+2z-3\phi)\;\;.
\eq
The singularity, parametrized by $z$, must be again of branch-point
type, and is situated at
\bq
\phi_0 = {1+2z\over3}\;\;\;,\;\;\;F_z(\phi_0) = {4\over27}(1-z)^3\;\;\;,\;\;\;
F''_z(\phi_0) = -2(1-z)\;\;,
\eq
so that the asymptotic behaviour for high powers of $\xi$ is given by
\bq
\phi_z(\xi) \sim \suml_n {1\over n^{3/2}\sqrt{27\pi}}
{(27/4)^n\over(1-z)^{3n-1}}\xi^n\;\;.
\eq
This immediately gives the behaviour with powers of $z$ as well.
Expanding the form $(1-z)^{-3n+1}$,
\bq
\phi(\xi,z) \sim \suml_{n,k}
\xi^nz^k{(27/4)^n\over\sqrt{27\pi}}{(3n-2+k)!\over
k!(3n-2)!n^{3/2}}\;\;,
\eq
gives for the number of graphs  
\bq
a(\f\to k\f,n\chi\bchi) \sim
{(27/4)^n\over\sqrt{27\pi}}{(3n-2+k)!n!^2\over(3n-2)!n^{3/2}}\;\;.
\eq
The goodness of this asymptotic result, when compared with the exact
enumeration, does not depend on $k$ but only on $n$: the exact result
is 1.3644356 times the approximate one for $n=1$, 
which ratio decreases to
1.0244771 for $n=10$ and to 1.0120180 for $n=20$. This $k$-independence 
is related to the fact that we have here the exact Taylor expansion
of the pole around $z=1$: when we use Stirling's approximation for the
factorials, a dependence on $k$ is introduced.\\

We may extend our discussion to the case of more fermions. If we introduce
$f$ fermion flavours, each flavour $j$ will have its own generating
functions $\psi_j$ and $\bpsi_j$, with variables $x_j$ and $\bx_j$: but
$\phi$ will still be described by \eqn{QEDeqn}, with the sole redefinition
\bq
\xi = x\bx \to \xi = \suml_{j=1}^fx_j\bx_j\;\;.
\eq
The amplitude therefore becomes, upon multinomial expansion:
\bqa
&&a(\f\to k\f,n_1\chi_1\bchi_1,n_2\chi_2\bchi_2,\ldots,n_f\chi_f\bchi_f)
\nl &&\;\;\;\;\;\;\; \sim
{(27/4)^n\over\sqrt{27\pi}}{(3n-2+k)!n!n_1!n_2!\cdots n_f!
\over(3n-2)!n^{3/2}}\;\;.
\eqa
For theories in which the various fermion types have the same type of vertices,
this is the general way in which one goes from the single-fermion to the
multi-fermion case, and therefore we shall only consider the
single-fermion case in the following.

The next simplest case is that where we allow a $\f^3$ self-interaction
for the boson in addition to the $\f\chi\bchi$ vertex. The analog of
\eqn{QEDeqn} is then
\bq
\phi = z + {1\over2}\phi^2 + {\xi\over(1-\phi)^2}\;\;,
\eq
so that
\bq
F_z(\phi) = (1-\phi)^2(\phi-{1\over2}\phi^2-z)
\eq
The equation $F'_z(\phi)$ now has two roots, leading however to the
same result: with $\phi=1-\Phi$ and $z=1/2-\zeta$,
the singularity condition reads $\Phi^2=\zeta$, and the
form of $\phi$ close to the singularity reads
\bq
\phi \sim \suml_n{2^{n-2}\over n^{3/2}\sqrt{\pi}}{1\over\zeta^{2n-1/2}}\;\;.
\eq
In turn, this gives the asymptotic form
\bq
a(\phi\to k\phi,n\chi\bchi) \sim
{2^{3n+k-5/2}\over n^{3/2}\sqrt{\pi}}{(2n+k-3/2)!n!^2\over(2n-3/2)!}\;\;.
\eq
Again, the goodness of the approximation is independent of $k$, and
the ratio exact/approximate reads 1.253314137 for $n=1$,
1.019251423 for $n=10$, and 1.009498692 for $n=20$.

\section{Asymptotics by saddle-point methods}
It is tempting to extend the reasoning of the previous section to
more complicated cases. This is, however, dangerous because of the following
reason. In finding the singularity we have first to solve $F'_z(\phi_0)=0$
to find $\phi_0$ as a function of $z$, and then to determine the coefficient
of $z^k$ in $1/F_z(\phi_0(z))^m$, where $m$ is a large number. 
In principle, this is then again determined
by the precise nature of the singularity of this form, that is, the
behaviour of $F_z(\phi_0(z))$ close to a root. If $z_0$ is this root, we
can generally write
\bq
F_z(\phi_0(z)) = A(z_0-z)^p\left( 1 + B(z_0-z) + {\cal O}\left((z_0-z)^2\right)
\right)\;\;,
\eq
for some positive $p$. Naively making the expansion around $z=z_0$ gives then
\bqa
&&{1\over F_z(\phi_0(z))^m} \sim
{1\over A^m(z_0-z)^{pm}} - {mB\over A^m(z_0-z)^{pm-1}}
+ \cdots\nl
&=& \suml_k {1\over A^mz_0^{pm}}
\left({z\over z_0}\right)^k\left[
{(pm+k-1)!\over k!(pm-1)!} - mBz_0{(pm+k-2)!\over k!(pm-2)!}+\cdots\right]\nl
&=& \suml_k {1\over A^mz_0^{pm}}\left({z\over z_0}\right)^k
{(pm+k-1)!\over k!(pm-1)!}\left[1
-Bz_0{m(pm-1)\over(pm+k-1)} +\cdots\right]\;\;.
\eqa
The correction term is {\em not\/} small for generic large $m$ and $k$ 
values, but only becomes small in the `super-asymptotic' limit $k\gg m^2\gg1$,
which is too asymptotic to interest us here.\\

The most efficient way to determine the high-$k$ behaviour appears to
be the following. Let us rewrite the relation between $\phi_0$ and $z$:
\bq
F'_z(\phi)=0 \Rightarrow G(\phi)=z\;\;.
\eq
For instance, in the `QED$+\f^3$' case considered above, we have
$G(\phi) = 2\phi-\phi^2-1/2$. The form of the coefficient $c_{n,k}$ of 
$\xi^nz^kn^{-3/2}$ is then given by a (counter-clockwise) 
Cauchy integral around $z=0$:
\bq
c_{n,k} \sim {1\over2i\pi}\oint\;dz\;
{1\over z^{k+1}}{1\over F_z(\phi)^n}
\sqrt{{-F_z(\phi)\over2\pi F''_z(\phi)}}\;\;\;,\;\;\;G(\phi)=z\;\;.
\eq
We can readily transform this into a loop integral for $\phi$:
\bqa
c_{n,k} &\sim& {1\over2i\pi}\oint\;d\phi\;
{G'(\phi)\over G(\phi)^{k+1}H(\phi)^n}
\sqrt{{-H(\phi)\over2\pi H_2(\phi)}}\;\;,\nl && H(\phi)=F_{G(\phi)}(\phi)\;\;,
H_2(\phi) =F''_{G(\phi)}(\phi)\;\;.
\eqa
Note that, in the definition of $H_2$, the substitution $z=G(\phi)$ must be
made {\em after\/} the differentiation.
The only tricky point is to determine which of the various roots of $G(\phi)=0$
should be chosen to integrate around. Note that this question {\em can\/}
be answered unambiguously: if worst comes to worst, one can simply
check the result against the exact enumeration for
largish values of $n$ and $k$.
Having found the point around which to do the $\phi$ integral, we then proceed
to deform the integration contour into a pair of integrals parallel to
the imaginary axis. The upwards integral (`main' integral) is chosen
to run over the saddle point $\hphi$, situated on the real axis and given by
\bq
K'(\hphi)=0\;\;\;,\;\;\;K(\phi) \equiv -n\log H(\phi) - k\log G(\phi)\;\;.
\eq
In many cases $\hphi$ can actually be given as a function of $n$ and
$k$ in closed form. The downwards integral (`return' integral) is chosen to
run over another saddle point. In the cases we have studied, we have always
found that the values of $H(\phi)$ and $F(\phi)$ in the saddle
point of the return integral are larger (in absolute value) 
than those for the saddle point of the main integral, so that the contribution
of the return integral is exponentially suppressed with respect to the
main integral. The result, therefore, is
\bq
c_{n,k} = {1\over2\pi}{G'(\hphi)\over G(\hphi)^{k+1}H(\hphi)^n}
\sqrt{{-H(\hphi)\over H_2(\hphi)K''(\hphi)}}\;\;.
\eq
The final asymptotic estimate for the number of tree graphs 
with $k+1$ external bosons and $2n$ fermions/antifermions can therefore
be written as follows:
\bqa
&&a(\f\to k\f,n\chi\bchi) \sim (n-1)!^2k!C_1^nC_2^{k+1}D\;\;,\nl
&&C_1 = 1/H(\hphi)\;\;,\nl
&&C_2 = 1/G(\hphi)\;\;,\nl
&&D = {n^{1/2}G'(\hphi)\over2\pi}
\sqrt{{-H(\hphi)\over H_2(\hphi)K''(\hphi)}}\;\;.
\eqa
The numbers $C_1$, $C_2$ and $D$ only depend on the ratio $k/n$.\\

As a first check, we redo the QED case with one fermion flavour. 
Here, $G(\phi)=(3\phi-1)/2$ so that we must integrate around $\phi=1/3$.
There is only a single saddle point $\hphi=(n+k)/(3n+k)$. Since
$\hphi>1/3$ and
main integral is indeed upwards, and the return integral
can be moved out to infinity. The result for $c_{n,k}$ is nothing
but the Stirling approximation of the `exact' result.\\

The next case is that of `QED$+\f^3$'. The equation $G(\phi)=0$ has two
roots, $1\pm\sqrt{1/2}$, of which $1-\sqrt{1/2}$ 
is on the appropriate Riemann sheet. This can be seen from the fact that
the saddle point $\hphi=(2n+k-\sqrt{2n^2+nk})/(2n+k)$ is to the
right of this root, and has a positive value for $G'(\hphi)$: choosing
the other possible saddle point (with a + sign before the square root)
leads to a negative $G'(\hphi)$ and consequently a negative asymptotic
estimate. Again, the saddle-point result boils down to the Stirling
approximation of the `exact' one.\\

A more complicated case is that of QED with a pure $\f^4$ interaction added.
We find
\bqa
G(\phi) &=& -{5\over12}\phi^3+{1\over4}\phi^2+{3\over2}\phi-{1\over2}\;\;,\nl
G'(\phi) &=& -{1\over4}(5\phi^2-2\phi-6)\;\;,\nl
H(\phi) &=& {1\over4}(1-\phi)^3(2-\phi^2)\;\;,\nl
H_2(\phi) &=& {1\over2}(1-\phi)(5\phi^2-2\phi-6)\;\;.
\eqa
Note that the roots of $H_2$ are also roots of $H$ or $G'$: this is 
a general occurrence.
The main-integral saddle point $\hphi$ is a root of the equation
\bq
(5n+3k)\phi^3-3(n+k)\phi^2-6(3n+k)\phi+6(n+k) = 0\;\;.
\eq
By standard methods, we can find the three roots of this equation:
\bqa
\hphi_r &=& {1\over5n+3k}\left[
2(31n^2+30nk+7k^2)^{1/2}\sin\left\{
{2r\pi\over3}-{1\over3}u\right\}
+n+k\right]\;\;,\nl
u &=&\arcsin\left(
{29n^3+75n^2k+63nk^2+17k^3\over(31n^2+30nk+7k^2)^{3/2}}\right)\;\;,
\eqa
with $r=0,1,2$. The choice $r=0$ interpolates smoothly between
.325259493 for $k/n=0$ and 1 for $n/k=0$, and this turns out to be the
correct saddle point. The singularity structure of the integrand is
as follows: there are poles of order $k+1$ at the three roots of $G$,
-1.788306912, .3252594930, and 2.063047419. There is a pole of order $3n-1$
at 1, and poles of order $n$ at $\pm\sqrt{2}$. With the standard convention
that the square root branch cut lies along negative real values, there
are cuts along the real $\phi$ axis from $-\sqrt{2}$ to
$(1-\sqrt{31})/5\sim-.9135528726$ and from $(1+\sqrt{31})/5\sim1.313552873$
to $\sqrt{2}$. The return integral can be taken to cross the real axis
at -.9135528726, where $H=2.041476117$ and $G=-1.344004635$. The
saddle-point values for the main integral are always smaller than these in
absolute value, confirming the above
statement that the return integral may safely be neglected.\\

The more relevant case of QCD is treated in the same manner. We have
\bqa
G(\phi) &=& -{5\over12}\phi^3-{3\over4}\phi^2+2\phi-{1\over2}\;\;,\nl
G'(\phi) &=& {1\over4}(4-5\phi)(2+\phi)\;\;,\nl
H(\phi) &=& {1\over4}(1-\phi)^3(2-2\phi-\phi^2)\;\;,\nl
H_2(\phi) &=& -{1\over2}(1-\phi)(4-5\phi)(2+\phi)\;\;,
\eqa
and $\hphi$ solves
\bq
(5n+3k)\phi^3+3(3n+k)\phi^2-12(2n+k)\phi+6(n+k) = 0\;\;;
\eq
this root can be written as
\bqa
\hphi &=& {1\over5n+3k}\left[
2(13k^2+49n^2+50nk)^{1/2}\sin\left({1\over3}u\right)-3n-k\right]\;\;,\nl
u &=& \arcsin\left(2{141n^3+225n^2k+123nk^2+23k^3\over
(13k^2+49n^2+50nk)^{3/2}}\right)\;\;.
\label{QCDsaddlepoint}
\eqa
The singularity structure resembles that of the previous case:
poles of order $k+1$ at the three values
-3.343142188, .2853836802, and 1.257758508; a pole of order $3n-1$ at 1,
and poles of order $n$ at $-1-\sqrt{3}\sim-2.732050808$ and
at $-1+\sqrt{3}\sim.7320508076$; and branch cuts running from
$-1-\sqrt{3}$ to -2 and from $-1+\sqrt{3}$ to 4/5. The loop-integration contour
is situated around .2853836802 and $\hphi$ moves smoothly from this
value upwards to .7320508076 as $k/n$ increases from 0 to infinity.
The saddle point for the return integral is at $\phi=-2$, where
$H=27/2$ and $G=-25/6$, again always considerably bigger than
$H(\hphi)$ and $G(\hphi)$.
This allows for the complete determination of $C_{1,2}$ and $D$:
the value of $\hphi$ is given by \eqn{QCDsaddlepoint}, and
\bqa
K''(\phi) &=& n{5\phi^4+12\phi^3+2\phi^2-36\phi+20\over
(2-2\phi-\phi^2)^2(1-\phi)^2}\nl &&
+3k{25\phi^4+60\phi^3+54\phi^2-204\phi+156\over
(5\phi^3+9\phi^2-24\phi+6)^2}\;\;.
\eqa
In the following table we present the non-universal quantities 
$C_1$, $C_2$ and $D$ for various ratios $k/n$.
\begin{center}
\begin{tabular}{|c|c|c|l|}
\hline\hline
$\log_{10}(k/n)$ & $C_1$ & $C_2$ & $D$ \\
\hline
-3.0   &8.143  &4155. &.0002929 \\
-2.5   &8.157  &1316. &.0005196 \\
-2.0   &8.217  &417.5 &.0009172 \\
-1.5   &8.389  &133.6 &.001594 \\
-1.0   &8.961  &43.78 &.002640 \\
-0.5   &10.89  &15.40 &.003802 \\
0      &18.00  &6.452 &.003882 \\
0.5    &48.97  &3.666 &.0002183 \\
1.0    &197.2  &2.833 &.0006782 \\
1.5    &836.8  &2.599 &.0001506 \\
2.0    &3141.  &2.535 &.00002913 \\
2.5    &10660. &2.516 &.000005336 \\
3.0    &34530. &2.511 &.0000009583 \\
\hline\hline
\end{tabular}
\end{center}
As expected, the accuracy of the asymptotic approximation improves uniformly
if $n$ and $k$ grow with a fixed ratio. In the table we collect
some results, where we have of course only those values for which
both $n$ and $k$ are integers, and have iterated the exact generating
function up to $n+k=29$. The accuracy is actually quite reasonable
even for moderate values of $n$ and $k$.
\begin{center}
\begin{tabular}{|c|c|c|c|c|c|c|c|}\hline\hline
$n$ & $k=n/3$ & $k=n/2$ & $k=2n/3$ & $k=n$ & $k=3n/2$ & $k=2n$ & $k=3n$ \\
\hline
1   &      &      &      &      &      &1.019 &1.003 \\
2   &      &.9904 &      &1.010 &1.010 &1.007 &.9995 \\
3   &.9709 &      &.9999 &1.006 &      &1.004 &.9993 \\
4   &      &.9927 &      &1.004 &1.004 &1.003 &.9994 \\
5   &      &      &      &1.003 &      &1.002 &.9995 \\
6   &.9837 &.9946 &.9992 &1.002 &1.003 &1.002 &.9995 \\
7   &      &      &      &1.002 &      &1.001 &.9996 \\
8   &      &.9958 &      &1.002 &1.002 &1.001 &      \\
9   &.9888 &      &.9993 &1.001 &      &1.001 &      \\
10  &      &.9966 &      &1.001 &1.001 &      &      \\
11  &      &      &      &1.001 &      &      &      \\
12  &.9915 &.9971 &.9994 &1.001 &      &      &      \\
13  &      &      &      &1.001 &      &      &      \\
14  &      &.9975 &      &1.001 &      &      &      \\
15  &.9932 &      &.9995 &      &      &      &      \\
16  &      &.9978 &      &      &      &      &      \\
17  &      &      &      &      &      &      &      \\
18  &.9943 &.9980 &      &      &      &      &      \\
19  &      &      &      &      &      &      &      \\
20  &      &      &      &      &      &      &      \\
21  &.9951 &      &      &      &      &      &      \\
\hline\hline
\end{tabular}
\end{center}

We may also consider a theory where $\f^p$ bosonic self-interactions
occur for every $p$. We simply list the results:
\bqa
G(\phi) &=& 3\phi-{1\over2}\phi\exp(\phi)-{1\over2}\exp(\phi)\;\;,\nl
G'(\phi) &=& 3-\exp(\phi)-{1\over2}\phi\exp(\phi)\;\;,\nl
H(\phi) &=& {1\over2}(-1+\phi)^3\left(-2+\exp(\phi)\right)\;\;,\nl
H_2(\phi) &=& -(-1+\phi)\left(-6+2\exp(\phi)+\phi\exp(\phi)\right)\;\;,\nl
0 &=& \exp(\hphi)((n+k)\hphi + n-k) - (6n+2k)\hphi + 2k\;\;,\nl
K''(\phi) &=& n{12-10\exp(\phi)+3\exp(2\phi)-4\phi\exp(\phi)+2\phi^2\exp(\phi)
\over(-2+\exp(\phi))^2(-1+\phi)^2}\nl
&&+ k{(6\phi\exp(\phi)+\exp(2\phi)+6\phi^2\exp(\phi)+36-24\exp(\phi))\over
(-6\phi+\phi\exp(\phi)+\exp(\phi))^2}\;\;.
\eqa
The Riemann sheet structure of the function $G$ is of course much more
complicated in this case, but fortunately the relevant saddle point
is the simplest solution on the real axis, again interpolating
smoothly between the appropriate zeroes of $H$ and $G$.
In the table, we give the non-universal quantities as a function of $k/n$.
\begin{center}
\begin{tabular}{|c|c|c|l|}
\hline\hline
$\log_{10}(k/n)$ & $C_1$ & $C_2$ & $D$ \\
\hline
-3.0 &8.150 &2000. &.0002904 \\
-2.5 &8.170 &1250. &.0005151 \\
-2.0 &8.224 &384.6 &.0009093 \\
-1.5 &8.396 &135.1 &.001580 \\
-1.0 &8.977 &44.05 &.002614 \\
-0.5 &10.91 &15.53 &.003753 \\
0    &17.99 &6.494 &.003797 \\
0.5  &48.08 &3.731 &.002092 \\
1.0  &184.5 &2.907 &.0006325 \\
1.5  &729.9 &2.681 &.0001372 \\
2.0  &2564. &2.618 &.00002613 \\
2.5  &8333. &2.591 &.000004754 \\
3.0  &20000 &2.584 &.0000008515 \\ \hline\hline
\end{tabular}
\end{center}
These numbers are qualitatively quite similar to that for QCD.
The accuracy of the approximation appears to be almost identical to
that of the QCD case.\\

Another interesting case is that of scalar QED, where we have an additional
$\f^2\chi\bchi$ interaction term. The SD equations now become more
complicated:
\bq
\psi = x + \psi(\phi+\phi^2/2)\;\;\;,\;\;\;
\bpsi = \bx + \bpsi(\phi+\phi^2/2)\;\;\;,\;\;\;
\phi = z + (1+\phi)\psi\bpsi\;\;,
\eq
so that
\bq
\phi = z + {\xi(1+\phi)\over(1-\phi-\phi^2/2)^2}\;\;.
\eq
Following the same steps as before, we find
\bqa
G(\phi) &=& {-2+9\phi^2+4\phi^3+6\phi\over3(2+2\phi+\phi^2)}\;\;,\nl
G'(\phi) &=& {4(\phi^2+2\phi+4)(1+\phi)^2\over3(2+2\phi+\phi^2)^2}\;\;,\nl
H(\phi) &=& {(2-2\phi-\phi^2)^3\over12(2+2\phi+\phi^2)}\;\;,\nl
H_2(\phi) &=& {(\phi^2+2\phi+4)(-2+2\phi+\phi^2)\over2+2\phi+\phi^2}\;\;;
\eqa
the main-integral saddle point is a solution of
\bq
(4n+k)\phi^3+3(3n+k)\phi^2+6n\phi-2(n+k) = 0\;\;,
\eq
and reads
\bqa
\hphi &=& {1\over4n+k}\left[
2(k^2+4nk+n^2)^{1/2}\sin\left({\pi+u\over3}\right)-3n-k\right]\;\;,\nl
u &=& \arcsin\left({n(3k^2+18nk+25n^2)\over(k^2+4nk+n^2)^{3/2}}\right)\;\;.
\eqa
Finally, we also need
\bqa
&&K''(\phi) = {4n(84\phi^2+26\phi^4+6\phi^5+\phi^6+24+56\phi+64\phi^3)\over
(-2+2\phi+\phi^2)^2(2+2\phi+\phi^2)^2} +\nl
&&{8k(36+320\phi^2+
371\phi^4+205\phi^5+72\phi^6+16\phi^7+2\phi^8+148\phi+432\phi^3)\over
(2+2\phi+\phi^2)^2(-2+9\phi^2+4\phi^3+6\phi)^2}.\nl
\eqa
The results for the non-universal constants are given in the following table.
\begin{center}
\begin{tabular}{|c|c|c|l|} \hline\hline
$\log_{10}(k/n)$ & $C_1$ & $C_2$ & $D$ \\ \hline
-3.0 &9.671 &4189. &.1731E-5 \\
-2.5 &9.690 &1326. &.5466E-5 \\
-2.0 &9.756 &420.2 &.1721E-4 \\
-1.5 &9.970 &133.7 &.5371E-4 \\
-1.0 &10.66 &43.18 &.1627E-3 \\
-0.5 &13.06 &14.55 &.4459E-3 \\
0    &23.03 &5.510 &.8568E-3 \\
0.5  &85.11 &2.667 &.6268E-3 \\
1.0  &805.8 &1.776 &.1574E-3 \\
1.5  &.1526E5 &1.496 &.2247E-4 \\
2.0  &.4021E6 &1.407 &.2542E-5 \\
2.5  &.1197E8 &1.379 &.2647E-6 \\
3.0  &.3712E9 &1.370 &.2681E-7 \\
\hline\hline
\end{tabular}
\end{center}
The non-universal constants appear to vary much more rapidly as a function
of $k/n$ than in the case of QCD. The accuracy of the asymptotic estimate, 
however, is essentially the same.

\section*{Conclusions}
We have demonstrated how the number of tree-level diagrams for several
theories can be computed exactly for given numbers of external legs
of various kinds. We have also described how asymptotic formulae for
these numbers, valid in the limit of many legs of each kind, can be
obtained. We have compared these results for several theories of interest,
including QED, QCD and scalar QED. Comparison with the exact results up
to fairly high order shows that the asymptotic estimates are accurate.


\begin{thebibliography}{9999}
\bibitem{BerGie} 
F.~Berends, W.~Giele, H.~Kuijf, Nucl. Phys. {\bf B321} (1989) 30;
Nucl. Phys. {\bf B333} (1990) 120

\bibitem{CarMor} 
F.~Caravaglios, M.~Moretti, 
Phys. Lett. {\bf B358} (1995) 332.

\bibitem{ArgKlePap}
E.N.~Argyres, R.~Kleiss, C.G.~Papadopoulos, Nucl.Phys. {\bf B391} (1993) 42

\bibitem{Kui}
J.G.M.~Kuijf, { \it Multiparton production at hadron colliders }, PhD thesis 
(University of Leiden) (1991)

\bibitem{Vol} 
M.B.~Voloshin, Nucl.Phys. {\bf B383}(1992) 233

\bibitem{Manetal}
F.~Caravaglios, M.L.~Mangano, M.~Moretti, R.~Pittau, 
Nucl.Phys. {\bf B539} (1999) 215

\end{thebibliography}
\end{document}